\documentstyle[aps,prl,epsfig,twocolumn]{revtex}
\begin{document}
\twocolumn[
\hsize\textwidth\columnwidth\hsize\csname
@twocolumnfalse\endcsname
\draft
\title{		Probing possible decoherence effects 
		in atmospheric neutrino oscillations}
\author{	E.\ Lisi$^a$, A.\ Marrone$^a$, and D.\ Montanino$^b$ }
\address{	$^a$Dipartimento di Fisica and Sezione INFN di Bari,
             	   Via Amendola 173, I-70126 Bari, Italy \\ }
\address{$^b$Dipartimento di Scienza dei Materiali dell'Universit\`a di Lecce,
		Via Arnesano,  I-73100 Lecce, Italy}
\date{\today}
\maketitle
\begin{abstract}
It is shown  that the results of the Super-Kamiokande atmospheric neutrino
experiment, interpreted in terms of $\nu_\mu\leftrightarrow\nu_\tau$  flavor
transitions, can probe possible decoherence effects induced by new  physics
(e.g., by quantum gravity) with high sensitivity, supplementing current
laboratory  tests based on kaon oscillations and on neutron interferometry. By
varying the (unknown) energy dependence of such effects, one can either obtain
strong limits on their amplitude, or use them to find an unconventional
solution to the atmospheric $\nu$ anomaly based solely  on decoherence.
\end{abstract}
]

The Super-Kamiokande (SK) atmospheric neutrino experiment  has found convincing
evidence \cite{Evid} for the quantum-mechanical phenomenon of $\nu$ flavor
oscillations \cite{Pont} in the $\nu_\mu\leftrightarrow\nu_\tau$  channel. Such
evidence consistently emerges from different  SK data samples  (sub-GeV
leptons,   multi-GeV leptons, and upgoing muons  \cite{Samp}) as well as from
other atmospheric $\nu$ experiments  \cite{Othe}.

The simplest model for $\nu_\mu\leftrightarrow\nu_\tau$ oscillations involves 
two neutrino states $\nu_1=(1,0)^T$ and $\nu_2=(0,1)^T$ with masses $m_1$ and
$m_2$, and two  flavor states $\nu_\mu=(c_\theta,s_\theta)^T$ and 
$\nu_\tau=(-s_\theta,c_\theta)^T$, where $\theta$ is the neutrino mixing angle,
$c=\cos$, $s=\sin$, and $T$ denotes the transpose.  The Liouville  equation for
the $\nu$ density matrix $\rho$,
\begin{equation}
			\dot\rho=-i[H,\rho]\ ,
\label{liouville}
\end{equation}
is then governed (in the mass basis) by the Hamiltonian 
		$H=\textstyle\frac{1}{2}\,{\rm diag}(-k,+k)$,
where  $k=\Delta m^2/2E$,  $\Delta m^2=m^2_2-m^2_1$, and $E\;(\gg m_{1,2})$  
is the $\nu$ energy (in natural units).  The solution $\rho(t)$ of
Eq.~(\ref{liouville}), with initial conditions $\rho(0)=\Pi_{\nu_\mu}$ (where 
$\Pi_{\nu_\mu}=\nu_\mu\otimes\nu_\mu^\dagger$ is the $\nu_\mu$ state
projector), gives the $\nu_\mu$ survival probability  after a length $x(\simeq
t)$,
\begin{equation}
	\textstyle
	P(\nu_\mu\leftrightarrow\nu_\mu)=
	{\rm Tr}[\Pi_{\nu_\mu}\rho(t)]=
	1-\frac{1}{2}s^2_{2\theta}(1-\cos kx)\ ,
\label{pimumu}
\end{equation}
which is the well-known oscillation formula \cite{Pont}.

Equation~(\ref{pimumu}) beautifully fits the SK data \cite{52kt} over a wide
range of $\nu$ energies ($E\sim 10^{-1}$--$10^3$ GeV) and flight lengths 
($x\sim 10^1$--$10^4$ km), provided that $\Delta m^2\simeq 3\times10^{-3}$
eV$^2$ and $s^2_{ 2\theta}\simeq 1$ \cite{52kt,3atm}.  Such striking agreement
severely constrains possible deviations from the standard hamiltonian $H$ 
\cite{3atm,Lipa}.    In this work we show that the SK data can also be used to
probe deviations from the standard Liouville dynamics in Eq.~(\ref{liouville}),
that might be induced by new physics beyond the  standard electroweak model.

In general,  modifications of Eq.~(\ref{liouville}) emerge from dissipative
interactions with an environment \cite{book}, and   can be parametrized by
introducing  an extra term $\cal D[\rho]$,
\begin{equation}
		\dot\rho=-i[H,\rho] - {\cal D}[\rho]\ ,
\label{liouvilleD}
\end{equation}
which violates the conservation of ${\rm Tr}(\rho^2)$ and allows transitions
from pure to mixed states.  The operator $\cal D$ has the dimension of an
energy, and its inverse defines the typical (coherence) length after which the
system gets mixed \cite{Conv}.

Among the possible sources of decoherence, a particularly intriguing   one
might be provided by quantum gravity, as  suggested by Hawking in the context
of black-hole thermodynamics \cite{Ha75}. From such a viewpoint, any physical
system is inherently ``open,'' due to its unavoidable, decohering interactions
with a pervasive ``environment''  (the spacetime and its Planck-scale  dynamics
\cite{Foam}).  Following the pioneering paper \cite{El84},  quantum gravity
decoherence effects have been investigated in oscillating systems which
propagate over macroscopic distances (see \cite{Erev} for reviews). Analyses
have been mainly focused on $K\overline K$ oscillations  \cite{El84,Pesk,kaon}
and on neutron interferometry \cite{El84,neut}, by assuming reasonable
phenomenological forms for  $\cal D$. In both systems, no evidence has been
found for  ${\cal D}\neq 0$, and strong limits have  been derived on the
quantities parametrizing $\cal D$ \cite{Erev}:
\begin{equation}
		||{\cal D}|| \lesssim 10^{-21}{\rm \ GeV}\ 
		(K\overline K,\ n\ {\rm systems})\ .
\label{Dlimits}
\end{equation}

Theoretical estimates for $||{\cal D}||$ are very uncertain \cite{El84}, and
can range from unobservably small values up to the limits in  (\ref{Dlimits}).
Therefore, it is wise to adopt a phenomenological viewpoint, trying to learn
from experiments and to improve the laboratory limits (\ref{Dlimits}) with
novel approaches, such as those provided by $\nu$ oscillations. Indeed,
attempts have been made to explain the solar $\nu$ deficit through decoherence
\cite{nuap,sola,atmo}. It has also been suggested that decoherence might play a
role in interpreting the atmospheric $\nu$ data \cite{atmo,Gros} although,  to
our knowledge, no detailed analysis of the SK results has been attempted so
far. The crucial point is that, for typical atmospheric $\nu$ energies  ($
10^{0\pm1}$ GeV),  the oscillation length $\lambda=2\pi/k$ spans the range
$\sim 10^{3\pm1}$ km; then,  if the (de)coherence length  $\ell$  is of
comparable size, terms as small as $||D||\sim \ell^{-1}  \sim 10^{-22\pm1}$ GeV
can be probed.

In order to fix a well-defined framework,  we specialize Eq.~(\ref{liouvilleD})
under reasonable (although not compelling) phenomenological assumptions. The
most general requirement is perhaps that of {\em complete positivity\/}
\cite{CPos,Bena},  corresponding to assume a linear, Markovian, and
trace-preserving map $\rho(0)\to\rho(t)$. This implies the so-called Lindblad
form \cite{Lind} for the decoherence term,
\begin{equation}
		{\cal D}[\rho] = \textstyle\sum_n 
		\{\rho,D_nD_n^\dagger\} -2D_n\rho D^\dagger_n \ ,
\label{lindblad}
\end{equation}
where the operators $D_n$ arise from tracing away the environment dynamics (see
\cite{Lrec} for a recent proof). Master equations of the Lindblad form are
ubiquitous in physics (see \cite{book,Acca} for theorems and applications).  
Concerning $\nu$ oscillations, such equations describe $\nu$ propagation in
dissipative media as, e.g., matter with fluctuating density \cite{fluc} or
thermal baths \cite{stod}.  Here, however, the environment embeds possible new
physics (e.g., the spacetime ``foam'' \cite{Foam}) for which there is no
established theory.

In the absence of first-principle calculations, we assume that at least the
laws of thermodynamics hold in the $\nu$ system. The time increase of the von
Neumann  entropy $S(\rho)=-{\rm Tr}(\rho \ln \rho)$ can be enforced by taking
$D_n=D^\dagger_n$ \cite{entr}, so that Eq.~(\ref{lindblad}) becomes  ${\cal D}[
\rho ]= \sum_n [D_n,[D_n,\rho] ]$. The conservation of the average value of the
energy [${\rm Tr}(\rho H)$] requires, in addition, that $[H,D_n]=0$ 
\cite{Pesk,Jliu}.

The hermitian operators $\rho$, $\Pi_{\nu_\mu}$, $H$, and $D_n$, can be 
expanded \cite{book} onto the  basis formed by the unit matrix $\bbox{1}$ and
by the Pauli matrix vector  $\bbox{\sigma}=(\sigma_1,\sigma_2,\sigma_3)^T$. We
take
${\rho} 	= \frac{1}{2}(\bbox{1}+\bbox{p}\cdot\bbox{\sigma})$,
${\Pi_{\nu_\mu}}= \frac{1}{2}(\bbox{1}+\bbox{q}\cdot\bbox{\sigma})$,
${H} 		= \frac{1}{2}\bbox{k}\cdot\bbox{\sigma}$, and
${D_n} 		= \frac{1}{2}\bbox{d}_n\cdot\bbox{\sigma}$,
where 
$\bbox{q}=(s_{2\theta},0,c_{2\theta})^T$ and $\bbox{k}=(0,0,-k)^T$.
Defining 
$ {G}=\sum_n |\bbox{d}_n|^2\bbox{1} - \bbox{d}_n\otimes \bbox{d}_n^T$,
Eq.~(\ref{liouvilleD}) is transformed into a Bloch equation, 
	${\bbox{\dot{p}}}=\bbox{k}\times\bbox{p}-{G}\cdot\bbox{p}$,
which  has a simple physical interpretation: the standard term 
$\bbox{k}\times\bbox{p}$ induces $\nu$ oscillations, while the decoherence term
${G}\cdot\bbox{p}$ is responsible for their damping \cite{book,stod}.

The requirement  $[H,D_n]=0$ implies that each vector $\bbox{d}_n$ is parallel
to $\bbox{k}$ \cite{Jliu}. Therefore, the tensor ${G}$ takes  the form
${G}={\rm diag}(\gamma,\gamma,0)$ with $\gamma=\sum_n|\bbox{d}_n|^2\geq0$
\cite{matr}.  The general solution $[\bbox{p}(t)={V}\cdot \bbox{p}(0)]$ of the
Bloch equation is then given by the evolution operator
\begin{equation}
{V}=\left(
\begin{array}{ccc}
+ e^{-\gamma t}\cos kt 	& + e^{-\gamma t}\sin kt 	& 0 \\
- e^{-\gamma t}\sin kt 	& + e^{-\gamma t}\cos kt 	& 0 \\
  0 			&          0 		 	& 1
\end{array}
\right)\ .
\end{equation}

If the system is prepared in the pure (zero entropy)  $\nu_\mu$ state
[$\bbox{p}(0)=\bbox{q}]$, the asymptotic final state   is
$\bbox{p}(\infty)=(0,0,c_{2\theta})$. Since ${\rm
Tr}[\rho^2(\infty)]=(1+c^2_{2\theta})/2 < 1$ and  $S[\rho(\infty)]=-c^2_\theta
\ln c^2_\theta - s^2_\theta\ln s^2_\theta > 0$, the system evolves indeed into
a  mixed state with positive entropy.  Maximal entropy  $(S=\ln 2)$ corresponds
to maximal $\nu$ mixing $(s^2_{2\theta}=1)$. Purity and entropy are conserved
only if $\rho$ is  prepared in a pure mass eigenstate $[\bbox{p}(0)=(0,0,\pm
1)^T]$.

The survival probability  $P_{\mu\mu}=\frac{1}{2}(1+\bbox{q}^T\cdot V \cdot
\bbox{q})$ reads
\begin{equation}
\textstyle P_{\mu\mu} 	= 1-\frac{1}{2}s^2_{2\theta}
		(1-e^{-\gamma x}\cos kx)\ ,
\label{pgamma}
\end{equation}
which reduces to the standard expression (\ref{pimumu}) in the limit $\gamma\to
0$. For $\gamma x \sim O(1)$, one expects significant deviations from the usual
oscillation fit to the SK data.

We make a quantitative study of the effects of $P_{\mu\mu}$ in (\ref{pgamma}),
by computing the theoretical SK lepton distributions in zenith angle
$(\vartheta)$, and by fitting them to the SK data through a $\chi^2$
statistics, as extensively discussed in \cite{3atm}. The main difference from
\cite{3atm} is:  (i) the 30 data bins for the SK distributions  refer to a
longer detector exposure (52 kton$\cdot$year \cite{52kt}); (ii) the oscillation
probability is here taken from Eq.~(\ref{pgamma}). In the fit, we study both
the case with $(\Delta m^2,s^2_{2\theta},\gamma)$ unconstrained (oscillations
plus decoherence) and the case with $\Delta m^2=0$ and ($s^2_{2\theta},\gamma)$
unconstrained (decoherence only). We find significant differences in the
results, depending on the energy variation assumed for $\gamma$ (which is not
necessarily a constant parameter). For definiteness, we discuss only three
scenarios, corresponding to a possible power-law dependence of the kind $\gamma
= \gamma_0 (E/{\rm GeV})^n$ with $n=0$, 2, and $-1$.

For $n=0$ ($\gamma=\gamma_0=\rm const$) the best fit with oscillations  plus
decoherence ($\chi^2_{\min}=22.6$) is reached  for $\Delta m^2=3\times 10^{-3}$
eV$^2$, $s^2_{2\theta}=1$, {\em and\/}  $\gamma_0=0$, which corresponds to the 
case of pure   $\nu_\mu\leftrightarrow\nu_\tau$ oscillations. Since no evidence
is seen to emerge for decoherence effects, meaningful upper bounds on the
parameter $\gamma$ can be placed. By taking  $\chi^2-\chi^2_{\min}=6.25$
(corresponding to 90\% C.L.\ for three degrees of freedom), we get
\begin{equation}
	\gamma_0 < 3.5\times 10^{-23}{\rm\  GeV\ \ } (n=0)\ .
\label{gamma}
\end{equation}
The limits at 95\% and 99\% C.L.\ are found to be $4.1\times 10^{-23}$ GeV and
$5.5\times 10^{-23}$ GeV, respectively. The bound (\ref{gamma}) shows  that:
$(i)$ if decoherence effects have the same origin (e.g., quantum gravity)  and
similar size in the different  $K$, $n$, and $\nu$ systems, then atmospheric
$\nu$ observations can improve the current laboratory limits (\ref{Dlimits});
and  $(ii)$ decoherence effects, if any, can develop only over a typical length
scale $\ell = \gamma_0^{-1} \gtrsim 5600$~km.

\begin{figure}[t]
\vspace*{-3.6truecm}
\hspace*{-1.2truecm}
\epsfig{%
bbllx=1.4truecm,bblly=1.3truecm,bburx=19.5truecm,bbury=26.5truecm,%
height=13.truecm,figure=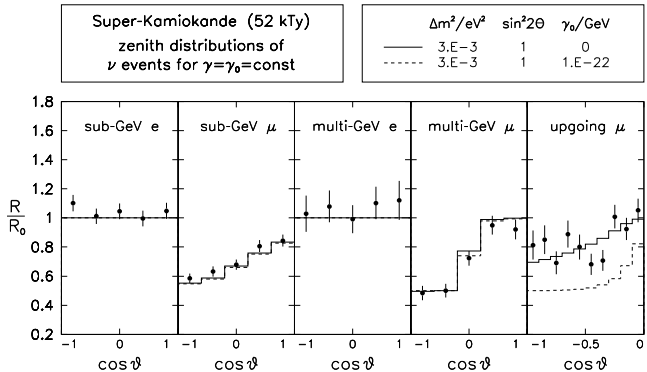}
\vspace*{-4.6truecm}
\caption{Effects of decoherence $(\gamma_0\neq0)$ on the distributions
	of lepton events as a function of the zenith angle ($\vartheta$). The
	SK data are shown as dots with $\pm 1\sigma$ error bars. The histograms
	represent our theoretical calculations. In each bin, the electron $(e)$
	and muon $(\mu)$  rates $R$ are normalized to standard (no oscillation,
	no decoherence) expectations $R_0$.}
\label{fig1}
\end{figure}

Figure~\ref{fig1} shows (for $n=0$)  the zenith distributions of SK events for
best-fit standard oscillations ($\gamma_0=0$) and in the presence of an
additional decoherence term ($\gamma_0=10^{-22}$ GeV).  The electron $(e)$
distributions  are unaffected  $(P_{ee}=1)$.  In the sub-GeV $\mu$ sample,
decoherence is almost unobservable, due to the large intrinsic smearing
\cite{3atm} of both energy and angle. In the multi-GeV $\mu$ sample,  the
transition from no oscillation ($P_{\mu\mu}\sim 1$ for $\cos\vartheta\sim 1$)
to averaged oscillations  ($P_{\mu\mu}\sim 1/2$ for $\cos\vartheta\sim -1$) is
made only slightly faster by decoherence effects. Such effects are instead
dominant in  the higher-energy sample of upgoing $\mu$, where the oscillation
phase $kx$ is small, and decoherence generates a much faster suppression of
vertical muons  $(\cos\vartheta\sim -1)$, corresponding to the longest $\nu$
flight lengths. Finally, we find a bad fit ($\chi^2\gtrsim 49$) when
oscillations are switched off [$H=0$, corresponding to $k=0$ in
Eq.~(\ref{pgamma})].  Therefore, in the case $n=0$, the SK data cannot be
explained solely by decoherence.

The case $n=2$ may also be of phenomenological interest, in the light of a 
possible dimensional guess of  the form  $\gamma\propto E^2/M_P$ \cite{Pred}. 
In this case,  decoherence effects are even more disfavored than for $n=0$,
since they  produce a faster suppression of muons with increasing energy,
contrary to observations. We find an upper limit  $\gamma_0 < 0.9\times
10^{-27}$ GeV at  90\% C.L.\  [to be compared with the limit (\ref{gamma})].
For $k=0$ (decoherence without oscillations) the fit is also very bad
($\chi^2\gtrsim 70$).

From the previous cases ($n=0$ and $n=2$) we learn that decoherence effects can
be strongly constrained, the more the faster they increase with energy.
Conversely, we expect weaker constraints for a {\em decreasing\/} energy
dependence,  such as for $\gamma\propto E^{-1}$ $(n=-1)$.

The case $n=-1$ may also be motivated by assuming that the exponent  in
Eq.~(\ref{pgamma}) behaves as a Lorentz scalar. A boost from the $\nu$ rest
frame to the laboratory frame would then introduce a factor $m_\nu/E$ (just as
for the oscillation phase),  giving a decoherence parameter of the form
$\gamma=\gamma_0(E/{\rm GeV})^{-1}$. Of course, this ansatz should be taken
with a grain of salt, since dissipative equations  are known to entail problems
with Lorentz invariance \cite{Pesk,Sred}  (however, see \cite{Jliu,Unru}). In
any case, assuming $\gamma=\gamma_0(E/{\rm GeV})^{-1}$,  we have performed a
fit to the SK data with  $(\Delta m^2,s^2_{2\theta}, \gamma_0)$ unconstrained.
The best fit is reached, once again, for $\gamma_0=0$, but the upper limit on
$\gamma_0$ is now relatively weak, $\gamma_0<2\times 10^{-21}$~GeV at 90\% C.L.
Therefore, for $n=-1$, one may add sizable decoherence effects to oscillations,
without destroying the agreement with SK data. Can one switch off completely
oscillations, and explain the data as a pure decoherence effect? The answer is
surprisingly positive.  For $\Delta m^2=0$, the best agreement with the data 
is reached at $s^2_{2 \theta}=1$ and $\gamma_0=1.2 \times 10^{-21}$~GeV, with 
$\chi^2_{\min}/N_{\rm DF}=27.1/(30-2)$, giving a good fit. This case represents
a novel solution to the atmospheric $\nu$ anomaly, based solely on decoherence.

\begin{figure}[t]
\vspace*{-3.6truecm}
\hspace*{-1.2truecm}
\epsfig{%
bbllx=1.4truecm,bblly=1.3truecm,bburx=19.5truecm,bbury=26.5truecm,%
height=13.truecm,figure=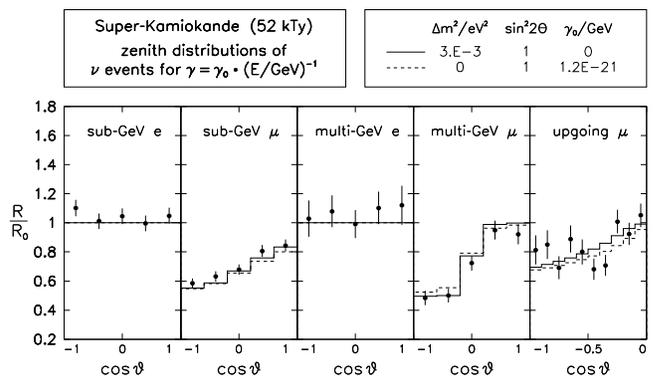}
\vspace*{-4.6truecm}
\caption{Comparison of best-fit scenarios for pure oscillations (solid line, as
	in Fig.~\protect\ref{fig1}) and for pure decoherence with 
	$\gamma\propto 1/E$ (dashed line).}
\label{fig2}
\end{figure}

Figure~\ref{fig2} show such ``exotic'' best fit (decoherence without
oscillations)   as compared to the ``canonical'' best fit (oscillations
without  decoherence). The two cases appear to be almost indistinguishable
within errors, although they entail completely different physics. It is amusing
to notice that, for the two best-fit cases of Fig.~\ref{fig2}, the $\nu_\mu$
survival probability approximately read
\begin{eqnarray}
P_{\mu\mu} &\simeq& \textstyle \frac{1}{2}[1+\cos(+\beta\cdot L/E)]\ \ \ 
 ({\rm pure\ oscillations}),\\
P_{\mu\mu} &\simeq& \textstyle \frac{1}{2}[1+\exp(-\beta\cdot L/E)]\ \ \ 
({\rm pure\ decoherence}),
\end{eqnarray}
where $E$ is in GeV, $L$ is the $\nu$ pathlength (km), and $\beta \sim 7\times
10^{-3}$ GeV/km. Both cases have the same asymptotic behavior, namely, $\langle
P_{\mu\mu}\rangle\simeq 1$ ($\frac{1}{2}$) for small (large) $L/E$. For
intermediate values of $L/E$, the strong differences between the oscillating
cosine factor and the monotonic exponential damping appear to be effectively
suppressed by the large smearing in the $\nu$ energy and angle, due to the
interaction and detection processes in SK. Therefore, future long-baseline
accelerator experiments (such as K2K, MINOS, and the CERN-to-Gran Sasso project
\cite{DiLe}) will be crucial in discriminating the above two functional forms
for $P_{\mu\mu}$, by revealing the oscillation (or damping) pattern now hidden
by smearing effects.

Finally, we test the best-fit decoherence case of Fig.~\ref{fig2} against the
negative results of current $\nu_\mu\to\nu_\tau$  appearance searches
\cite{DiLe}. In the CHORUS and NOMAD experiments \cite{Shor} one has $\langle
L/E\rangle \simeq 0.025$~km/GeV  and $P_{\mu\tau}=1-P_{\mu\mu} \simeq
\frac{1}{2}\beta \langle L/E \rangle$  (for $\Delta m^2=0$ and
$s^2_{2\theta}=1$).  Then the experimental limit $P_{\mu\tau}\lesssim 1.3\times
10^{-4}$  \cite{DiLe} implies the upper bound $\beta \lesssim 1.1\times
10^{-2}$~GeV/km, which is compatible with the best-fit value $\beta\sim7\times
10^{-3}$ GeV/km.

In conclusion, we have performed a phenomenological analysis of modifications
of  the Liouville dynamics,  in the context of  atmospheric
$\nu_\mu\leftrightarrow\nu_\tau$  transitions. Within a  simple model embedding
the relevant physics (oscillations plus decoherence), we have found that the
Super-Kamiokande data can be a sensitive probe of decoherence effects (e.g.,
originated by quantum gravity), supplementing current laboratory tests based on
$K$ and $n$ interferometry. Depending on the energy behavior assumed for such
effects, one can either constrain them strongly, or use them to explain the
atmospheric $\nu$ anomaly without oscillations.

We thank J.~Ellis, R.~Floreanini, S.\ Pakvasa, and S.~Pascazio for useful
comments.


{\em Note added.\/} After submission of this {\em Letter\/},  two related works
appeared \cite{BF00}. We also noted a recent preprint \cite{Ad00}  suggesting
an exceedingly small theoretical estimate for $\gamma$ ($\sim k^2/M_P$), which
would discourage current experimental tests with neutrinos (as well as with
kaons and neutrons). It seems to us that such estimate \cite{Ad00}, being
essentially based on a dimensional guess, should be presently considered with
great caution. In the absence of both a full dynamical theory and of {\em ab
initio\/} calculations for decoherence effects,  any current ansatz may prove
to be wrong. This fact warrants phenomenological analyses as ours, whose
results, inferred from experimental data,  remain valid independently of
(uncertain) guesses about  the origin and the size of $\gamma$.




\begin{thebibliography}{99}
\bibitem{Conv} 	In this context, it is useful to remind the conversion constant
		$(1 {\rm\ km})^{-1}=1.97\times 10^{-19}$~GeV.
\end{thebibliography}
\end{document}